\documentclass[mathleft
]{an}
\usepackage{graphicx}
 \usepackage{amsmath,amssymb,amsfonts,eucal,mathrsfs} 
\usepackage{times}
\usepackage{tabularx}
\begin{document}

% The following seven commands are intended for editorial usage and should be ignored by
% the author(s).
\Pagespan{789}{}% Document's page range.
% If second parameter is left empty, the last page is computed automatically.
\Yearpublication{2009}%
\Yearsubmission{2008}%
\Month{11}%
\Volume{999}%
\Issue{88}%
% \DOI{This.is/not.aDOI}%

\title{The FIRST radio survey: The $K-z$ diagram of FIRST radio
  sources identified in the Bo\"{o}tes and Cetus fields}

\author{K. EL Bouchefry\fnmsep\thanks{
  \email{kelbouchefry@gmail.com}\newline}
%Example
%for footnote, note the usage of the \texttt{fnmsep}
%command as separator between institute number and footnote mark}
%\and  Others\inst{2,3}
}

\titlerunning{The Hubble diagram}
\authorrunning{ EL Bouchefry  2009}
\institute{Astrophysics and Cosmology Research Unit, University of KwaZulu-Natal, Westville, 4000, South Africa}
%\and
%Downing Street 10, London, UK
%\and
%The second affiliation of the second author}

\received{2008 Dec 12} \accepted{\textit{:revised}}
\publonline{*******}


\abstract{%
  This paper presents the Hubble diagram ($K-z$ relation) for FIRST (Faint Images of the Radio 
Sky at 20 cm) radio sources identified in the Bo\"{o}tes and Cetus fields. The  correlation between the $K$ magnitude of the FIRST-NDWFS sample  and the photometric redshifts found to be linear. The dispersion about the best fit line is given by 1.53 for the whole sample and 0.75 at $z>1$. The paper also presents a composite   $K-z$ diagram of  FIRST radio sources and low-frequency selected radio samples with   progressively fainter flux-density limits (3CRR, 6C, 7CRS and the EIS-NVSS sample). The majority of FIRST radio sources lie   fainter than the no evolution curve ($3\,L_{\star}$ galaxies) probably highlighting the fact  that the galaxy luminosity is correlated with the radio power.}
 \maketitle

\section{Introduction}

Powerful radio sources have played a crucial role in our understanding of galaxy evolution. The host galaxies of powerful radio sources, such as radio galaxies and quasars, are identified with giant elliptical (Best et al. 1998, Mclure \& Dunlop 2000, Jarvis et al. 2001, De Breuck et al. 2002, Willott  et al. 2003, Zirm et al. 2004) and are associated with the most massive black holes (Taylor et al. 1996, Dunlop et al. 2003, Mclure et al. 2004, Mclure \& Jarvis 2004) in the universe. Studies of these objects at high redshift have shown a tight correlation in the Hubble $K-z$ diagram for  powerful radio sources. The infrared \textit{K}-magnitude against redshift relation has been widely used as a tool for investigating the evolution with cosmic epoch of stellar populations of luminous galaxies, since K corrections, dust extinction corrections, and the effect of any secondary star formation are all relatively unimportant at near infrared wavelengths. It has played an important infrared role in the search for and the study of high redshift galaxies (e.g Lilly \& Longair 1984, Eales et al. 1987). For example, the first radio galaxy discovered at $z>3$ was selected on the basis of a faint $K\,\sim 18.5$ magnitude (Lilly 1988). The Hubble $K-z$ diagram is known to be an excellent tool to measure stellar masses of galaxies up to higher redshift (Eales et al. 1999, van Breugel et al. 1998, Lacy et al. 2000) and has been widely used to study the evolution  in radio galaxies. 

Lilly \& Longair (1984) showed for the 3CRR sample that the $K-z$  relation is well defined with approximately constant dispersion to redshifts $z > 1$ and indicates evolution in the galaxies'luminosity of about a magnitude at $z\sim1$ if $\Omega_0=1$. They concluded that the giant elliptical hosting
the low redshift ($z<0.6$) radio galaxies from the 3CRR sample are the result of passively evolving stellar populations which formed  at high redshift (e.g Lilly 1989). Subsequent studies based on low frequency selected radio samples with successively fainter flux-density limits have been subject to
a degeneracy between radio luminosity and redshift (see e.g.; 6CE sample which is 5 times fainter than the 3CRR sample (Eales et al. 1997), 7CIII, 7C, 20 times fainter than 3CRR sample (e.g. Lacy et al. 2000; Willott et al. 2003)).

Willott et al. (2003) have investigated the $K-z$ relation for 205 radio galaxies with high spectroscopic completeness ($ z\sim 0.05-4.4$) obtained from a combined data set selected at different flux limits; 3CRR (Laing, Riley \& Longair 1983), 6CE (Eales et al. 1997, Rawlings, Eales \& Lacy 2001), 6$C^{\star}$ (Jarvis et al 2001a,b) and  the 7CRS (Lacy et al. 2000, Willott et al. 2003) showing that 3CRR and 7CRS radio galaxies offset by $\sim 0.55 $ in the \textit{K}-magnitudes over all redshift while the 6C differ from the 3CRR ones by $\sim0.3$ mag.  These results have been interpreted as a correlation of both properties with black whole mass (Willott et al. 2003, McLure et al. 2004).  The best fit for the combined sample (3CRR, 6C\footnote{The 6C sample refers to the 6CE and the  6C$^{\star}$samples}, 7CRS) quoted by Willott et al. (2003)  is: $ K(z)=17.37 + 4.53\, \log_{10}\,z -0.31\, (\log_{10}\,z)^{2} $. The brightest sample is 3CRR selected at 178 MHz with a flux density limit of $S_{178} \geq 10.9$ Jy, the intermediate samples are the 6CE and 6C* selected at 151 MHz with flux density limits of $2.0 \leq S_{151}\leq 3.93$ Jy and $0.96 \leq S_{151}\leq 2.00$ Jy respectively. The 7CRS selected at 151 MHz with flux density limits of $S_{151} \geq 0.5$ Jy.

In order to overcome the drawbacks of using the most powerful radio galaxies, and flux-limited samples, it is important to select samples that cover a wide range of flux-density limits and redshift. This was one of the primary motivation for developing a combined EIS-NVSS radio sample which is 12 times fainter than 7CRS survey (CENSORS: Best et al 2003). Brookes et al. 2006 established  a $K-z$ relation for their radio galaxies of the CENSORS and used it to calculate redshift for non spectroscopically identified sources in their sample (Brookes et al. 2008).  The EIS-NVSS sample (Brookes et al. 2006) has been selected at 1.4 GHz with flux density limit of 7 mJy.

In  EL Bouchefry (2008a), the author defined the FIRST-Bo\"{o}tes/Cetus  radio sample and presented robust optical /infrared counterparts to FIRST  radio sources. Based on the multi-wavelength (\textit{Bw R I J K}), photometric redshift has  been calculated using the public code \textit{Hyperz}. In EL  Bouchefry (2008b), the optical/infrared properties of FIRST radio sources  identified in Bo\"{o}tes/Cetus fields and their host environment is  discussed. This paper shed light on the $K-z$  relation of the FIRST radio sources identified  in the Bo\"{o}tes and  the Cetus fields ($33^{\circ} \leq \delta \leq 36^{\circ}$, $216^{\circ} < \alpha \leq 221^{\circ}$ ). These data are combined  with those from the 3CRR, 6CE  (Rawlings et al. 2001) 6C*  (Jarvis et al. 2001a,b) and 7C-III (Lacy et  al. 2000) and EIS-NVSS (Brookes  et al. 2006) to define the $K-z$ relation  over 400 radio galaxies ranging  from 1Jy to 1 mJy levels. Section 2 describes the radio and optical data. Section  3 presents the Hubble diagram of  the FIRST-Bo\"{o}tes/Cetus radio sample,  and  conclusions  are summarised in section 4. 

Throughout this paper it is  assumed that $H_{\circ}=70~{\rm km~s^{-1}~Mpc^{-1}}$, $\Omega_{M} =0.3$, and $\Omega_{\Lambda} = 0.7$ unless
stated otherwise.

\section{The sample data}

\subsection{The NDWFS survey}

The NOAO Deep Wide Field Survey (NDWFS) is a deep multi-band imaging (\textit{Bw, R, I, J, H, K}) designed
to study the formation and evolution of large scale structures (Jannuzi et al. 1999; Brown et al. 2003). This survey consists of two fields\footnote{http://www.noao.edu/noao/noaodeep/}; the first one  is 
located in the  Bo\"{o}tes field centred on approximately $\alpha = 14^{h} \; 30^{'}\; 05.7120^"$, $\delta = +34^{\circ} 16^{'} 47.496^{"}$, covering a 3 by  3 square degrees region, and the latter one is located in a 2.3 by 4 square degrees region in the Cetus field.  The survey catalogue has been split by declination range into four strips ($32^{\circ}\leq \delta <33^{\circ}, 33^{\circ} \leq \delta <34^{\circ}, 34^{\circ} \leq \delta <35^{\circ}, 35^{\circ}\leq \delta <36^{\circ}$); each strip observed in four bands (\textit{Bw, R, I, K}). Only the last two strips has been considered in a  previous study of FIRST radio sources in the Bo\"{o}tes field (EL Bouchefry \& Cress 2007). The magnitude limits are: $Bw\sim 25.5$ mag, $R\sim25.8$ mag, $I\sim25.5$ mag and
$K\sim19.4$ mag.

\subsection{The FLAMINGOS survey} 

 FLAMEX (Elston et al. 2006) is a wide area, deep near infrared imaging survey
 that covers 7.1  square degrees within the NDWFS survey regions; 4.7
 square degrees in the Bo\"{o}tes  field and 2.4 square degrees in the Cetus field in both \textit{J} and \textit{K} filters. The  FLAMEX catalogue is publicly  available
 \footnote{http://flamingos.astro.ufl.edu/extragalactic/overview.html}. This survey has been used to get infrared data for the second strip  ($33^{\circ}  \leq \delta < 34^{\circ}$) of the Bo\"{o}tes field.
\subsection{The FIRST catalogue}
The radio data are from the 2002 version of the  FIRST (Faint Images of the Radio Sky at Twenty-Centimetres; Becker et al 1995) Very Large Array catalogue\footnote{The FIRST catalogue is available online at http://sundog.stsci.edu}, and it is  derived from 1993 through 2002 observations. The FIRST  radio survey  has been carried out in recent years   with the VLA in its B-configuration to produce a map of 20 cm (1.4 GHz)  sky with  a beam size of 5.4 arcsec and an rms sensitivity of about 0.15 mJy/beam. The 2002 version of the catalogue covers a total of about 9033 square degrees of the sky (8422 square degrees in the north Galactic cap and  611 square degrees in the south Galactic cap); and contains 811,117 sources from the north and south Galactic caps. The
accuracy of the radio position depends on the brightness and size of the source and the noise in the map. Point sources at the detection limit of the catalogue have positions accurate to better
than 1 arcsec at $90\%$ confidence; 2 mJy  point sources typically have positions good to 0.5 arcsec. The radio surface density is $\sim 90$ deg$^{-2}$. About 900 sources fall within the  Bo\"{o}tes field and $\sim 242$ radio sources fall within the Cetus field.  

\section{The Hubble diagram}

The total number of FIRST radio sources identified in the Bo\"{o}tes field is 688/900 radio sources (either in one or more bands). There are 338 ($48\%$)  FIRST radio sources optically identified in $K$ band, and 273 ($39\%$)  were identified in four bands. In Cetus field, there are 113/242 ($47\%$) counterparts to FIRST radio sources in $J$ band, 124/242 ($51\%$) candidates in $K$ band. For a simple morphological classification, I used the Source Extractor stellarity parameter S/G (Bertin \& Arnouts 1996) provided by the NDWFS catalogue.  The Sextractor parameter has values between 0 (galaxy, more to the point, non-star)  and 1 (for point-like sources; QSOs or stars). All the sources with ${\rm S/G} < 0.7$ were classified as galaxies and sources with ${\rm S/G} >0.7$ were classified as point-like objects (QSOs or stars). The classification yielded  235 ($34\%$) radio sources classified as galaxy (identified in four bands) and 33 ($5\%$) radio sources classified as point-like objects. For sources identified in $K$ band, 261 ($37\%$) were classified as galaxies and 77 (11 \%) as point-like objects. The point-like objects tend to have a blue colour which is consistent with those of QSOs, so they are likely QSOs. A total number of 22 sources that were classified as point-like sources are  spectroscopically identified in the SDSS and were confirmed to be QSOs.

\subsection{The $K-z$ diagram for FIRST radio sources}

\begin{figure}
\begin{center}
\begin{tabular}{c}
\includegraphics{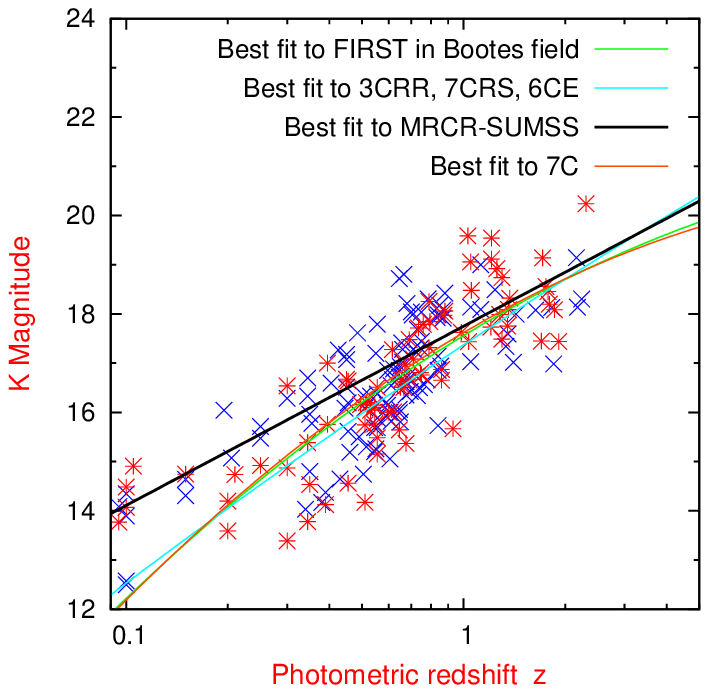} \\ %K_zlf3336.eps} \\
\includegraphics{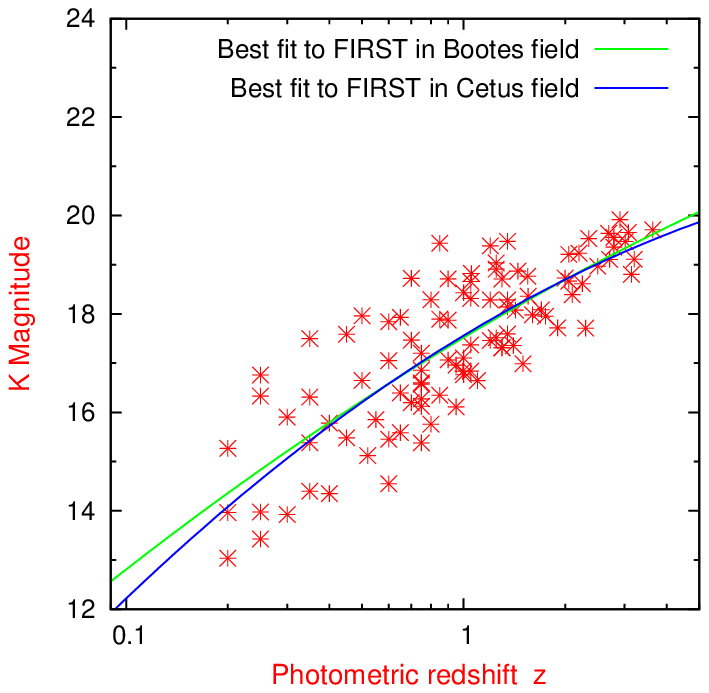} \\ % cetuskz.eps}  \\
\end{tabular}
\caption{\textit{top panel}: \textit{K} band magnitude against redshift for all FIRST radio
sources identified in Bo\"{o}tes field.  Crosses denote sources identified in the range  $34^{\circ} \leq \delta <36^{\circ}$  (introduced in EL Bouchefry \& cress 2007) and stars indicate the
 new counterparts of FIRST radio sources in the second strip  ($ 33^{\circ} \leq \delta <34^{\circ})$. The lines show the best fitting relationship between \textit{K} magnitude
  and $\log_{10}(z)$ for different samples. \textit{lower panel}: \textit{K} band magnitude against redshift for all FIRST radio sources identified in Cetus field.} \label{kz3336}
\end{center}
\end{figure}

In order to investigate the $K-z$ relation for the FIRST-Bo\"{o}tes/Cetus radio
sample, I used the best photometric redshift estimates with $\chi^{2}< 2.7$ ($\%90$ confidence) for all FIRST radio sources identified in \textit{Bw, R, I, K} (Bo\"{o}tes field) and \textit{I, J} (Cetus field). The point-like  sources were excluded from the $K-z$ analysis due to the fact that the $K-z$ relation is applicable only to radio galaxies for which the $K$ band emission is dominated by an old elliptical galaxy.

The $K-z$ relation for the new counterparts of FIRST radio sources in the second strip ($33^{\circ} \leq\delta<34^{\circ}$) augmented with those introduced in EL Bouchefry \& Cress 2007, in Bo\"{o}tes field, represented by stars and crosses respectively, is shown in Figure \ref{kz3336} (top panel).  The green line is the best fit second order polynomial relationship between the \textit{K} band and $\log_{10}\,z$ for all the data (in Bo\"{o}tes field):

\begin{eqnarray}
  \label{eq:kzbootes}
  K(z)=17.56 + 4.14\, \log_{10}\,z -1.20\, (\log_{10}\,z)^{2},
\end{eqnarray}

\noindent the cyan line is the best fit of the  combined sample of 3CRR, 6C and 7CRS
(Willott et al. 2003), the black line illustrates the best fit of Bryant et al. (2009), and the red line shows the best fit for the 7CRS sample
alone. Brookes et al. (2006) claim that this latter is similar to the fit they
obtained for the CENSORS survey. Compared to the best fit of Willott et al. (2003), the fit for the FIRST radio sources is shifted slightly to fainter magnitude ($\sim 0.15\, {\rm mag}$) which could support finding that brighter radio sources are associated with galaxies that are
brighter in \textit{K} band, even for faint radio sources. More complete and deeper
sample is required to investigate this further. A subsample of FIRST radio
 sources with flux-densities greater than 10 mJy is considered but found no
better fit to the Willott et al. (2003) relation.\\

 Figure \ref{kz3336} (lower panel) displays the \textit{K}  band  magnitude
of FIRST radio sources against photometric redshift obtained for sources identified in Cetus field. The blue line in the figure shows the best fitting to the  $K-z$ relation for FIRST in Cetus field. The best fitting  to  the $K-z$ relation (in Cetus field) is:
\begin{eqnarray}
  \label{eq:kzcetus}
  K(z)=17.52 + 4.09\, \log_{10}\,z -0.62\, (\log_{10}\,z)^{2},
\end{eqnarray}

\noindent and the best fit for all the data (Bo\"{o}tes and Cetus field) is:

\begin{eqnarray}
  \label{eq:kztotal}
  K(z)=17.50	 + 4.13\, \log_{10}\,z -0.30\, (\log_{10}\,z)^{2}.
\end{eqnarray}

Recently, Bryant et al. (2009) have compiled a sample of 234 ultra-steep spectrum (USS) selected radio sources in order to find high redshifts sources. These authors have spectroscopic information  for only 36 sources ($15\%$). Bryant et et al. (2009) have investigated the $K-z$ diagram and quoted three fits (see Table 1).  Their fit to the $K-z$ relation was found to be fainter than the Willott et al. (2003) by more than 0.3 magnitudes at all redshifts.  However,  complete sample, deep infrared data and spectroscopic information are required in order to well understand the $K-z$ relation as this diagram has been widely used to calculate redshift in the absence of spectroscopic information.\\

\subsection{Dispersion in the $K-z$ relation}
The increase in the dispersion about the $K-z$ relation has been used to study the evolution  of  the stellar population in powerful galaxies, and to probe the  formation epoch of radio galaxies. For example, Eales et al. (1997) have concluded that the radio galaxies are in the formation epoch at $z=2$ based on their study to the sample B2/6C, while at $z<2$ the radio galaxies are passively evolving. They found that  the dispersion in the $K-z$ relation at high redshifts ($z>2$) of the B2/6C sample is 2.7 times greater  at low redshift ($z<2$).  Using the 7CIII sample, Lacy et al. (2000) found similar effect. Jarvis et al. (2002), found no evidence of an increase in the dispersion, and concluded that most radio galaxies have formed their stars at $z>2.5$ and passively evolved since then based on their study to the 6C sample. Willott et al. (2003) also found no increase in the dispersion about the $K-z$ relation in agreement with Jarvis et al. (2001). Bryant et al. (2009), have also calculated the dispersion  about the best-fit to $K-z$ relation as a function of redshift. These authors found a standard deviation $\sigma=0.7$ that is approximately constant at all redshifts (see their table 4), supporting the results found by Jarvis et al. (2001) and Willott et al. (2003) that radio galaxies hosts have been evolving passively since epochs corresponding to  $z=3$ or earlier. In this study, the dispersion about the mean $K-z$ relation is given by $\sigma_K= 1.53$. Similar correlation is found in other bands (\textit{Bw, R} and \textit{ I}) but with a larger scatter. One notes that the scatter in \textit{K} band is smaller at high redshift ($\sigma_K =0.76$ at $z>1$),  than at lower redshift, consistent with the idea that fainter radio survey probably include fainter galaxies that are different from the typical galaxies associated with bright radio sources. For the combined 3CRR, 7CRS, 6C data of Willott et al. (2003), $\sigma=0.58$ at redshifts up to $z=3$. A reason for the increased spread at lower radio luminosities is that a wider range  of galaxy masses host these radio sources. This could be related to the different radio emission mechanisms (most FIRST radio sources have FRI luminosities, where as most of the 3CRR, 6C, 7CRS sample are FRII radio sources). Owen \& Laing (1989) found that FRIs reside in hosts which on average 0.5 magnitudes brighter than those of FRII of comparable radio luminosity. Perhaps FRIs also have a broader spread in host luminosity. But the question is how much the photometric redshift errors contribute to the increased scatter in the $K-z$ relation. It is well known that the accuracy of photometric redshift increases with the number of bands used to calculate redshift (especially infrared bands). Here, only sources identified in four or five bands  were considered for the $K-z$ relation (in order to get accurate photometric redshifts). The total number of sources identified in four bands is governed by the lower fraction of sources identified in \textit{ K} band, which means only bright sources were  included, and there could be a considerable number of sources fainter ($K>20$) than the magnitude limit of the \textit{K} band data. With the additional criterion of $\chi^2 < 2.7$ ($90\%$ confidence limit) one excludes sources with a not very good fit and this makes the sample even more incomplete. These issues could also contribute to the flatness of the $K-z$ slope and the large scatter in the $K-z$ relation.\\

 One should also note that,  the Bo\"{o}tes and Cetus samples are not complete. As
 mentioned earlier, the Bo\"{o}tes
 field is split by declination into four strips; $32^{\circ}<\delta\leq 33^{\circ}$,  $33^{\circ}<\delta\leq 34^{\circ}$,  $34^{\circ}<\delta\leq 35^{\circ}$,  $35^{\circ}<\delta\leq 36^{\circ}$. No infrared data is availabe for the
 first strip, and the other strips are partially covered in $K$ band. Moreover, the $K$ band catalogue is not as deep as the 
\textit{Bw, R} and \textit{I} catalogues. There is only a fraction of sources identified in $K$ band compared to the other bands.  The Bo\"{o}tes and Cetus $K$ band magnitude histograms drop of at 19.5 mag. Therefore, the lack of $K>20$ could cause several effects:

\begin{enumerate}

\item  The slope of the $K-z$ relation for FIRST radio is a little bit flatter than for  the combined sample 3CRR, 6C, 7CRS.\\
%\item  Scatter in FIRST is quoted as 1.53 for whole sample and 0.76 and $z>1$. This difference can be explained if most of the missing galaxies are $K>20$ and $z>1$.\\
\item There could be a significant number of $K$ band faint radio galaxies missing (due to the incompleteness of the optical survey), and because of this most of the sources will lie at the bright end of $K$ band. Therefore, the observed FIRST $K-z$ relation having a very small offset from  the combined sample 3CRR, 6C, 7CRS of 0.15 mag at $z=1$ could be misleading and in reality there could be a larger difference in magnitudes between the samples.\\
\end{enumerate}

\begin{figure}
\begin{center}
\includegraphics[width=80mm]{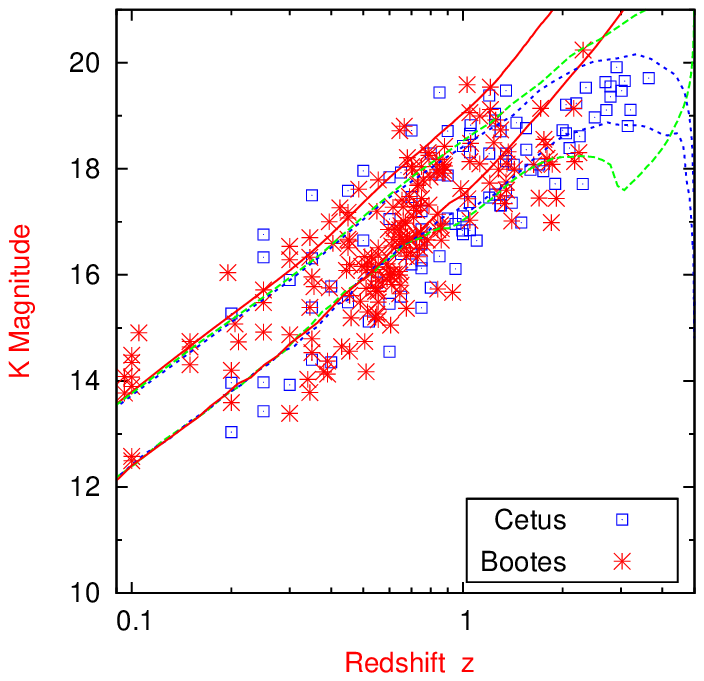}%{kzbootesEVO.eps}
\caption{The $K-z$ Hubble diagram for all FIRST radio sources identified in
 Bo\"{o}tes and Cetus fields. The three upper curves show: a non
evolution curve (red colour), instantaneous starburst ($0.1~{\rm Gyr}$) beginning at $z=5$ (blue colour) and a starburst (green colour) lasting $1~{\rm Gyr}$ and starting at $z=5$ (Jarvis et al. 2001). The three lower curves are: non evolving galaxies
  with  luminosity $1\,L_{\star}$ (red curve), the blue and green curves correspond 
to a passive evolution for galaxies with  $1\,L_{\star}$ assuming all stars in the galaxies 
formed at $z=5$ and $z=10$ respectively.} \label{kzbootesevo}
\end{center}
\end{figure}

\begin{figure*}
  \begin{center}
    \begin{tabular}{c}
      \resizebox{135mm}{!}{\includegraphics{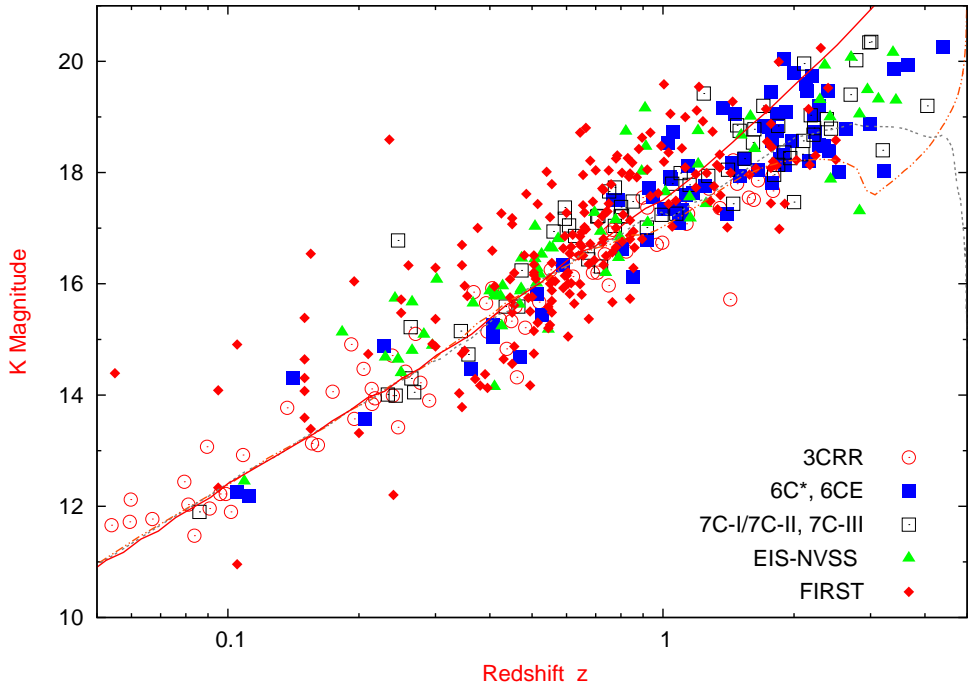} }%alK_zsamples.eps
    \end{tabular}
   \caption{The $K-z$ Hubble diagram for radio galaxies for the 3CRR, 6CE, 6C*,
     7CI/7CII, 7CIII, EIS-NVSS and NDWFS-FIRST-FLAMEX samples. Over-plotted are: a non
evolution curve (red colour), instantaneous starburst (black colour)
($0.1~{\rm Gyr}$) beginning at $z=5$ (orange colour) and a starburst
lasting $1~{\rm Gyr}$  and starting at $z=5$  (Jarvis et al. (2001)).}
\label{kz_combined}
  \end{center}
\end{figure*}

\begin{figure}
\begin{center}
\includegraphics{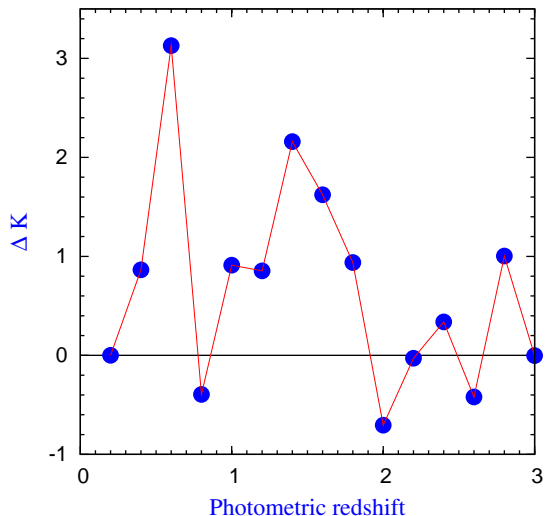}%{kzbootesEVO.eps}
\caption{The offset, $\Delta K$, of the FIRST radio sources compared to the best fitting $K-z$ relation for radio galaxies from Willott et al. (2003). Here $\Delta K= K- K_{fit}$. The data are plotted in steps of $\Delta z=0.2$.} \label{delta_fit}
\end{center}
\end{figure}

\begin{table*}
%\begin{small} \end{small}\begin{scriptsize}
\begin{footnotesize}

\begin{center}
\caption{Different fits to the $K-z$ relation for different samples}
 \begin{tabular}{lccll}
  \hline\hline
Sample & Flux limit &  Frequency &  Best fit to $K-z$ & References  \\
\hline
\hline
    &        &            &          &                   \\
 3CRR & 10.9 ${\rm Jy}$ & 178 ${\rm MHz}$  &     &   \\

 6C & 2.0 ${\rm Jy}$ & 151 ${\rm MHz}$ &   $K(z)=17.37 + 4.53 \,\log_{10} \,z -0.31(\log_{10} z)^{2}$    & Willott et al. (2003)   \\

7CRS & 0.5 ${\rm Jy}$ & 151 ${\rm MHz}$  &    &  \\

                \\
 NVSS-EIS    &  7.2 ${\rm mJy}$ & 1.4 ${\rm GHz}$ &  $K(z)=17.62 + 4.04 \,\log_{10} \,z -1.39(\log_{10} z)^{2}$           &   Brookes et al. (2006)                \\
\\
 FIRST-Bo\"{o}tes    &  1.0 ${\rm mJy}$&  1.4  ${\rm GHz}$ &   $K(z)=17.90 + 4.30 \,\log_{10} \,z$ & EL Bouchefry \& Cress 2007                   \\
\\
 FIRST-Bo\"{o}tes/Cetus    &  1.0 ${\rm mJy}$& 1.4 ${\rm GHz}$  &  $K(z)=17.50 + 4.13 \,\log_{10} \,z -0.30(\log_{10} z)^{2}$&    This work               \\

\\	

   & ---- & 874 ${\rm MHz}$ &  $K(z)=17.75 + 3.64 \,\log_{10}\,z$  at all redshift & Bryant et al. 
(2009)    \\
MRCR-SUMSS   & ---- & 874 ${\rm MHz}$ &  $K(z)=17.76 + 3.45 \,\log_{10}\,z$ at $z>0.6$& Bryant et al. (2009)    \\
 & ---- & 874 ${\rm MHz}$ &  $K(z)=17.89 + 3.11 \,\log_{10}\,z$ at $z>1$ & Bryant et al. (2009)    \\
     &        &            &          &                   \\
\hline
\hline
  
 \end{tabular}
\end{center}
\end{footnotesize}
%\end{scriptsize}
\end{table*}

In Figure \ref{kzbootesevo}, the three upper curves show non evolving and passive evolution $1\,L_{\star}$ galaxies as a function of redshift. The passive evolution models (blue and green plots) assume the stars in the galaxies formed at  $z=5$ and $z=10$ respectively. The three lower curves illustrate  passive stellar evolution tracks of a $3\,L_{\star}$ (K) galaxy for an instantaneous burst of star formation and one in which the burst of star-formation lasts  1 Gyr at $z=5$, as well as a no evolution curve as derived by Jarvis et al. (2001). The model curves in Figure \ref{kz_combined} show non evolving and passive evolution $3\,L_{\star}$. As clearly seen from the plots, the majority of the FIRST radio sources lie fainter than the no evolution curve ($3\,L_{\star}$ galaxies). This could be due to the fact that the  FIRST survey is deeper than the low-frequency complete samples that were used to make the $K-z$ diagrams in Jarvis et al. (2001)  and in Willott et al. (2003); probably highlighting a correlation between the galaxy luminosity and the radio power (Willott et al. 2003; McLure et al. 2004).

Figure \ref{kz_combined} shows  the near infrared Hubble diagram of\textit{ K} magnitude versus redshift $z$ for the FIRST-Bo\"{o}tes/Cetus sample combined with data of four samples compiled from the literature:  3CRR, 6CE, 6C*, and EIS-NVSS. The 3CRR, 6CE, 6C* and 7CRS have been gathered from the web site provided by Willott et al. (2003)\footnote{http://www.astro.physics.ox.ac.uk/~cjw/kz/kz.html}. The \textit{K}-band magnitudes and the corresponding redshift for the EIS-NVSS sample have been  compiled from the work of Brookes et al. (2006). All magnitudes are on the same metric system (64 kpc) except the FIRST-Bo\"{o}tes sample ($2''$ aperture).   The majority of the FIRST radio sources tend to be fainter than the other samples, this can be clearly seen   in Figure \ref{delta_fit} which shows the difference between the FIRST-Bo\"{o}tes/Cetus sample and the best fitting $K-z$ relation of Willott et al. (2003). This figure shows that the FIRST radio sources are fainter than the combined sample 3CRR, 6C, 7CRS over all redshifts, and things slightly change at redshifts greater than $\sim 2$. Brookes et al. (2006) found similar effects in their comparison of the EIS-NVSS sample to the best fitting $K-z$ relation of  Willott et al. (2003). Brookes et al. (2006) explained  that at these redshifts ($z>2$) there are no 3CRR sources and the best fit to the $K-z$ relation is a fit to the 6C and 7CRS samples alone (see their figure 3b).  However, the small fraction of the sources identified in $K$ band and the absence of spectroscopic information does not allow more conclusions.

\section{Conclusions}
In this paper, the $K-z$ diagram of the faint radio population detected at 1.4 GHz to a limiting flux density of 1 mJy has been investigated. The FIRST radio sources found to follow a similar $K-z$ relation to brighter radio samples, with evidence for a slight offset to fainter $K$ magnitude, consistent with the idea that faint radio survey include fainter galaxies that are different from the typical galaxies associated with bright radio sources. However, One should be aware of the limitation of the NDWFS data compared to the study of Willott et al. (2003). Willott et al. (2003) have obtained complete $K$ band data of the complete radio samples, while only a fraction of radio sources identified in the $K$ band in this study (due to the incompleteness of the NDWFS). Moreover, including only sources identified in several filter with the additional criterion of $\chi^2 < 2.7$ (accurate photometric redshift) make the sample more incomplete. These conclusions should be taken on the understanding that a high spectroscopic completness, deep K band data in the NDWFS survey, is required before one can draw any conclusions. But, it is encouraging that the $K$ band magnitudes with photometric redshift agree with the Willott et al. (2003) relation. 

This paper also presented a composite $K-z$ diagram of FIRST radio sources and
low-frequency selected radio samples with progressively fainter flux-density
limits: 3CRR, 6CE, 7CRS and the EIS-NVSS. The  majority of the FIRST radio
population tends to lie at fainter magnitudes possibly indicating that the
galaxy luminosity is correlated with radio power (Willott et al. 2003, Mclure
et al. 2004).

\acknowledgements
I would like to thank the anonymous referee for helpful
comments and suggestions which improved the paper.
I  also would like to  thank Dr Matt Jarvis and Prof Chris Willott for
Kindly supplying the stellar evolution curves. Thanks are also due to  Dr Antony Gonzalez  for answering all my questions concerning the FLAMEX surveys, and the South African Square Kilometre Array (SA SKA) project for  supporting  and funding my PhD studies. This work  makes  use of images data products provided by the NOAO Deep 
Wide-Field Survey (Jannuzi and Dey $1999$), which is supported by 
the National Optical Astronomy Observatory (NOAO). NOAO is
operated by AURA, Inc., under a cooperative agreement with the
National Science  Foundation.  This work also makes use of data products from
the FLAMEX survey. FLAMEX was designed and constructed by the
infrared instrumentation group (PI: R. Elston) at the University
of Florida, Department of Astronomy, with support from NSF grant
AST97-31180 and Kitt Peak National Observatory.

\end{document}